\documentclass[12pt]{iopart}
\usepackage{geometry}                
\usepackage[parfill]{parskip}    
\usepackage{graphicx}
\usepackage{amssymb}
\usepackage{multirow}
\usepackage[square, comma, sort&compress, numbers]{natbib}

\begin{document}

\title{Development of modularity in the neural activity of children's brains}

\author{Man Chen and Michael W. Deem\\
Department of Physics \& Astronomy\\
Department of Bioengineering\\
Center for Theoretical Biological Physics\\
Rice University\\
Houston, TX  77005}

\begin{abstract}
We study how modularity of the human brain changes
as children develop into adults.
Theory suggests that modularity can enhance 
the response function of a networked system subject to
changing external stimuli.
Thus, greater cognitive performance might be achieved
for more modular neural activity, and 
modularity might likely increase as children develop.
The value of modularity calculated from fMRI data is observed to increase
during childhood development and peak in young adulthood.
Head motion is deconvolved from the fMRI data, and it is shown that
the dependence of modularity on age is independent of
the magnitude of head motion.
A model is presented to illustrate how modularity can
provide greater cognitive performance at short times, i.e.\ task switching.
A fitness function is extracted
from the model. 
Quasispecies theory is used to predict how the average
modularity evolves with age, illustrating
the increase of modularity during development from children to adults
that arises from
selection for rapid cognitive function in young adults.
Experiments exploring the effect of modularity on
cognitive performance are suggested.
Modularity may be
a potential biomarker for injury, rehabilitation, or
disease.
\end{abstract}


\bibliographystyle{iopart-num}

\submitto{\PB}
\maketitle

\section{Introduction}

A modular organization of neural activity
can facilitate more rapid cognitive function, 
 because much of the rewiring of connections required for adaptation
is performed within the modules, which is easier and faster than
within the entire network
 \citep{Wagner,Lipson2002,Alon2005,Alon2007,jun,Dirk,Callahan2009}.
On the other hand, 
modularity may
restrict possible cognitive function, because a modular
neural architecture is a subset of all possible architectures
 \citep{Alon2005,Alon2007,jun,Dirk}.
Modularity in the neural activity of the human brain 
has been demonstrated \citep{Mountcastle,Fodor},
with activation of 
neural activity in different parts of the brain
observed by
functional magnetic resonance imaging (fMRI) 
\citep{Stanford,Schwarz,Ferrarini,Meunier2009,He,Chavez2010,Meunier2010}.
Remarkably, correlated neural activity 
can also be generated from free-streaming, subject-driven, cognitive states
 \citep{Stanford}.
%
Models of developing neural activity have shown a
self-organization of modular structure \citep{Rubinov2009}.


We here hypothesize that
selection for neural responsiveness is strong during childhood
and peaks during young adulthood.
Since modularity increases responsiveness,
we expect that modularity 
of neural activity in the brain might peak during adulthood as well.
These modules are correlated with physical 
structure of the brain \citep{Carlson2013}, but
they are not completely hard
coded at birth. Cognitive demands upon the brain
 promote development
of modular neural activity, which empowers the brain with
increased responsiveness and task switching ability.

A dynamic network of neural activity in the brain
that can reconfigure its architecture
will converge to a value of the modularity that depends on the
pressure upon it \citep{jun,Dirk}.
Here we show that modularity
of neural activity in the human brain increases from age 4 and
peaks in young adulthood. 
 We use a model to interpret these
results, showing that highly modular neural activity favors rapid,
low-level tasks, whereas less modular neural activity promotes
less rapid, effortful, high-level tasks.  The model shows that
increased connection between stored memories or faster required
response times for adults versus children can
explain the observed development of modularity in neural activity.

The structure of functional networks constructed from fMRI
data is age dependent. For example,
analysis of fMRI data from young adults and old adults shows
the modularity of human brain functional networks
decreases with age \citep{Meunier2009,Dirk,Onoda}.
Additionally, the architecture of the default network of the
brain extracted from fMRI data changes
as children develop into adults
\citep{Fair}, and  
working memory performance has been
shown to be related in an age-dependent way
with functional networks
\citep{Satterthwaite2013}.
That functional network constructed from fMRI data 
can be age dependent was emphasized in a study showing
subject age can be estimated from 5 minutes of fMRI  
data from individual subjects
\citep{Dosenbach}.
These works suggests a clear developmental and
age dependence of the networks constructed from fMRI data.
In other work, connection matrices averaged over ages were
constructed and the modularity 
of these average matrices was computed, observing
no trend of modularity with age \citep{Fair2009}.
The lack of observed trend in that study may be due to the construction of
averaged connection matrices, rather than consideration
of the connection matrices constructed from each individual.

Measurements of task switching costs, both mixing and switching,
show that young adults are more efficient than both children
and old adults at task switching \citep{Karbach}.
We suggest selection for neural task switching peaks during
young adulthood, and that selection for task switching is one
mechanism to explain the observed peak
of modularity in neural activity in young adulthood.
We demonstrate the cross-task utility of modularity in
a model of memory recall.  
In other words, in this model, we show
that modularity measured during the task of
watching a movie has functional implications for
a task such as memory recall.
Previous results show that regions of interest observed during
resting state activity strongly overlap with known functional regions
 \citep{Stanford,Carlson2013}.
Thus, it is likely that
the modularity observed during the task of movie watching has
functional implications for other tasks, such as memory recall.
Indeed, it has been demonstrated that modularity of
resting-state neural activity is positively
correlated with working memory capacity \citep{Stevens}.

A number of studies have reported that children move their heads
more than adults, and that this motion can affect properties computed
from the observed fMRI data.
Thus, it has been suggested that calculations should
separately account for head motion and age when drawing conclusions
about the effect of development on correlations of neural activity.
In some subject populations, with particular alignment methods,
spurious correlations
between the  computed network properties and extent of
head motion have been observed \citep{Satterthwaite}.  
Further work showed that the underlying age-dependent changes
to the spatial dependence of correlations persist after 
head motion correction, although the magnitude of the age dependence 
decreases \citep{Satterthwaite2013b}.
A study using the SPM2 software  showed
an apparent increase in local connectivity and decrease in
large-scale, distributed networks due to head motion \citep{Dijk}.
A study using in-house software  showed
an effect of head motion on extracted time series data \citep{Power}.  
A study using FSL
 showed a negative correlation between modularity
and head motion for motions above 0.07~mm \citep{Satterthwaite}.
 Since head motion is negatively
correlated with age,  
these studies caution that
artifacts due to head motion might 
erroneously be interpreted to predict that
modularity is positively correlated with age.
We quantify head motion artifacts, showing that
with the alignment procedure
and strict censor cutoff
used used in our study, effects of head
motion are negligible after alignment.


In this work, we analyzed
patterns of neural activity measured by fMRI for children
of different ages and for young adults watching a movie.
We show that a mathematical definition of modularity, using several different
basis sets, leads to the conclusion that modularity increases from childhood
to young adulthood.  These results are consistent with very
recent results showing increased within-module connectivity during
development \citep{Satterthwaite2013b}.
Theory suggests that highly modular neural
activity favors rapid, low-level tasks, whereas less modular neural
activity promotes less rapid, effortful, high-level tasks
 \citep{jun,Dirk,Meunier2010}.
We test the general predictions of this theory relating modularity
to performance and we derive a fitness function for a quasispecies
description of the modularity dynamics.
We describe details of the data analysis in the Methods section.
The Results section describes these results.
In the Model section, we introduce  a model to show that
a more modular neural architecture leads to more accurate
recognition of memory at short times.  We also show that
a more modular neural architecture leads to more accurate
recognition of memories when stored memories are more overlapping.
This model suggests that overlapping stored memories or faster
required response times for adults versus children are
mechanisms that could explain the
observed development of near-resting state modularity in neural activity.
We conclude with experimental and clinical implications
in the Discussion section.

\section{Methods}
\subsection{Subjects}

We analyzed fMRI data from 
21 adults age 18--26 and
24 children age 4--11 watching
20 minutes of Sesame Street \citep{Cantlon}.
Image acquisition details are provided in \citep{Cantlon}.
These data were taken from a set of 27 children (16 female) and
21 adults (13 female).  Three children were excluded due
to excessive head motion ($>5$ mm), opting-out, or experimenter error.
All children were typically developing.
In addition, screening for neurological abnormalities
was performed on all participants  \citep{Cantlon}.
Raw, KBIT-2 overall IQ scores are known for 17 of the children
and range from 18.5 to 66 \citep{Cantlon}.

\subsection{Processing of fMRI Data}

The two-dimensional fMRI data slices \citep{Cantlon} were combined
into a three-dimensional fMRI representation of the neural activity
as a function of time, with 2 sec time resolution and
4~mm spatial resolution.  The first six fMRI images of each
subject were discarded to allow the blood oxygen level dependent (BOLD)
signal to stabilize. 
The time series images for each subject were registered to
1~mm resolution, deskulled anatomical data, which created a
normalized image in standard space.
The time series images were then despiked, and the spike values were
interpolated using a non-linear interpolation \citep{afni}.
The images were then slice time corrected, registered, scaled into
Talarirach brain coordinates \citep{Talairach}, motion
corrected, bandpass filtered, and blurred (Cox, 1996).
Frame-to-frame motion of a subjects' head was corrected by
regressing out rigid translations and rotations of the fMRI data and
derivatives of these parameters during the time course.
Motion was censored with a threshold of 0.2,
i.e.\ RMSD of 2 mm or 2 degrees based on the Euclidean Norm
of the derivative of the translation and rotation parameters.
Outliers were censored with a threshold of 10\%. Due to
excessive censoring, 2 of the child and 1 of the adult subject data
were excluded.
For the individuals remaining in the study, a small fraction of the time
slices were removed by this censoring, also termed scrubbing, and
the values replaced with values linearly interpolated from neighboring time
points.  All calculations were performed with AFNI \citep{afni}.
The block of analysis code of example 9a in the afni\_proc.py was applied.

\subsection{Calculation of Modularity}

\begin{figure}[tb!]
\begin{center}
a) \includegraphics[width=0.4\columnwidth,clip=]{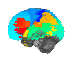}
b) \includegraphics[width=0.4\columnwidth,clip=]{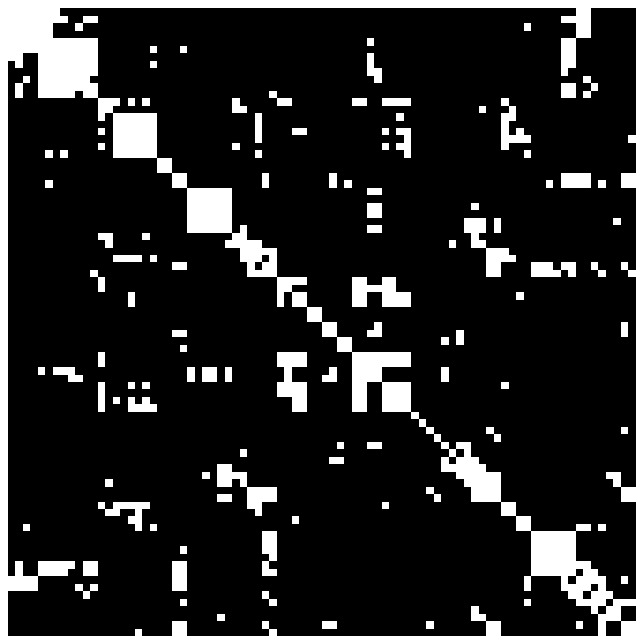}\\
c) \includegraphics[width=0.4\columnwidth,clip=]{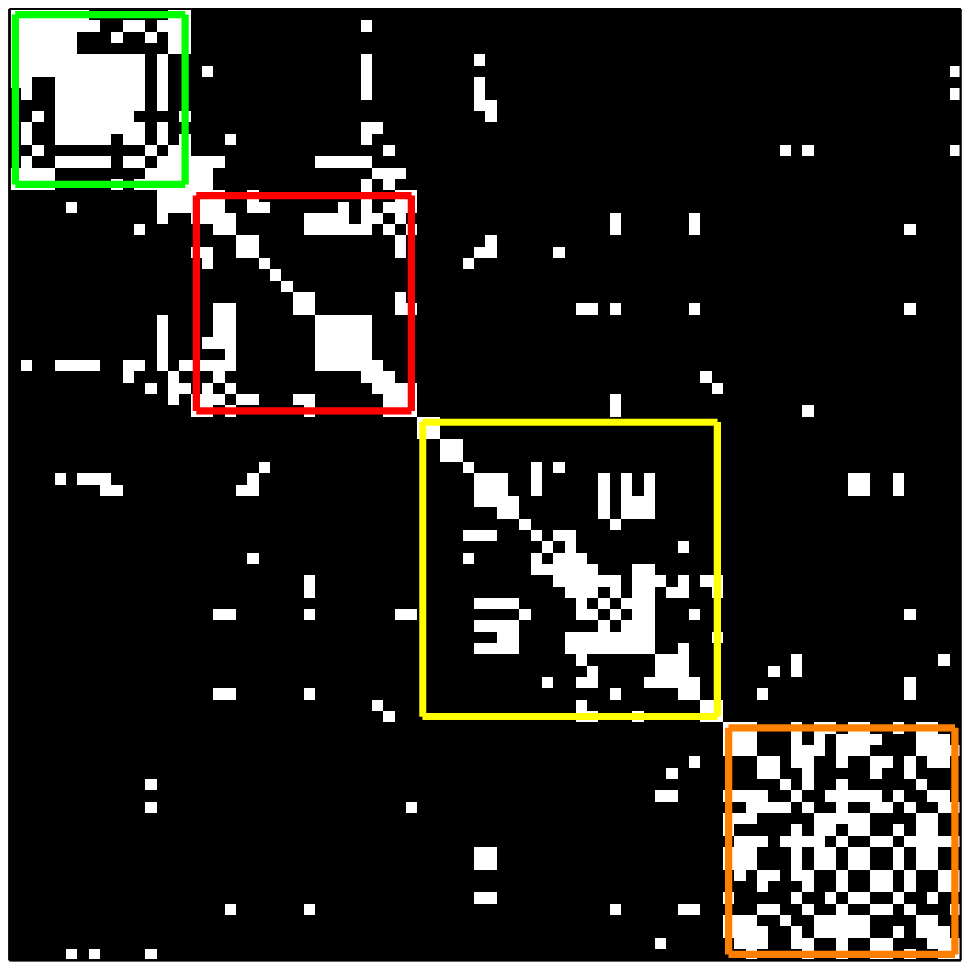}
d) \includegraphics[width=0.4\columnwidth,clip=]{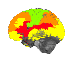}
\end{center}
\caption{Brain neural activity is clustered
into modules.
a) fMRI neural activity data are projected onto Brodmann areas,
shown as colored regions.
b) Neural activity between
different Brodmann areas is correlated
for subjects watching 20 minutes of Sesame Street, shown here
for one adult subject. Only the 
elements of the correlation matrix above a cutoff
are retained (white).
c) Modules are defined as the clusters that maximize
Newman's modularity.
The four modules identified from the 84 Brodmann areas
for this subject are shown,
of size
16 (green), 20 (red), 27 (yellow), and 21 (orange).
The modularity 
is 0.6441, with
contributions of 0.1585,   0.1310,  0.1592, and 0.1954 
from each module respectively.
d) The four modules of neural activity for this subject.
}
\label{fig1}
\end{figure}

We computed modularity from
correlations in neural activity of the brain
extracted from  fMRI data \citep{Schwarz,Meunier2009,Fortunato2010}.
Fig.\ \ref{fig1} illustrates the process.
Correlations were computed for each subject between Brodmann areas,
a standardized basis set describing regions of the human cerebral cortex
\cite{Nolte}.
Modularity was computed
using the Newman algorithm \citep{Newman2006}.
We projected the $N_{\rm edge}$ largest values of  the
correlation matrix to unity, and set the remaining values to zero.
We computed the modularity of this projected matrix.  These
values of modularity depend on the parameter $N_{\rm edge}$, and it will be confirmed
that adult modularity is greater than child modularity
for all $N_{\rm edge}$ values. The value for the $N_{\rm edge}$ parameter must be large enough
that the projected correlation matrix is fully connected, which 
implies $N_{\rm edge} \ge$ 200 for our data set. 
We considered 200 $\le N_{\rm edge} \le$ 500 so that
 the non-linear effect of the projection
were significant, i.e.\ the projected matrix was not simply all unity.

The numerical value of the
modularity is the probability to have
correlations within the modules, minus the probability
expected for a randomized matrix with the
same degree sequence \citep{Newman2006}.  In other words,
\begin{equation}
M = \frac{1}{2 e} \sum_{\rm all~modules} \sum_{\rm ~~~~areas~ i,j~ within~ this~ module}
\left(
A_{ij} - \frac{a_i a_j}{2 e}
\right)
\label{1}
\end{equation}
where $A_{ij}$ is one if there is an edge between 
Brodmann area $i$ and Brodmann area $j$ and
zero otherwise, 
The value of
$a_i = \sum_j A_{ij}$ is the degree of Brodmann area $i$, and $e = \frac{1}{2} 
\sum_i a_i$ is the total number of edges, as in Fig.\ \ref{fig1}.
Edges are established if the correlation between
Brodmann areas $i$ and $j$
is greater than the cutoff value, which is 
implicitly determined by the desired number of edges to keep in the
matrix, see Fig.\ \ref{fig1}b.
Modules are defined by the grouping that maximizes $M$
\citep{Newman2006}.
The Newman algorithm \citep{Newman2006} is used to
calculate values of modularity and the identity of the modules.
This algorithm gives a unique answer
for the modularity and set of modules
for a given connection matrix.
The value of the modularity depends on
how the Brodmann areas are grouped into modules,
 in Eq.\ \ref{1},
and the Brodmann areas are clustered into modules by
choosing the grouping that maximizes the modularity,
 Eq.\ \ref{1}.

It is not necessary to use a cutoff in the calculation of the modularity.
The Newman modularity calculation can be applied to a full matrix, 
with real number values, rather than the binary projection.
That is, the matrix $A$ in Eq.\ (\ref{1}) can simply
be the full correlation matrix, without projection.  
Modularity computations on the full, real-valued matrix
will also be presented.

Finally,  it is also not necessary to use the Brodmann areas as
regions of interest.  That is, the $i$ and $j$
in Eq.\ (\ref{1}) can simply be voxels, rather than Brodmann areas.
Results for (8~mm)$^3$ and (12~mm)$^3$ voxels will be presented.
Thus, modularity values for three different basis sets
were computed.

\section{Results}

\begin{figure}[tb!]
\begin{center}
a) \includegraphics[width= 0.45\columnwidth,clip=]{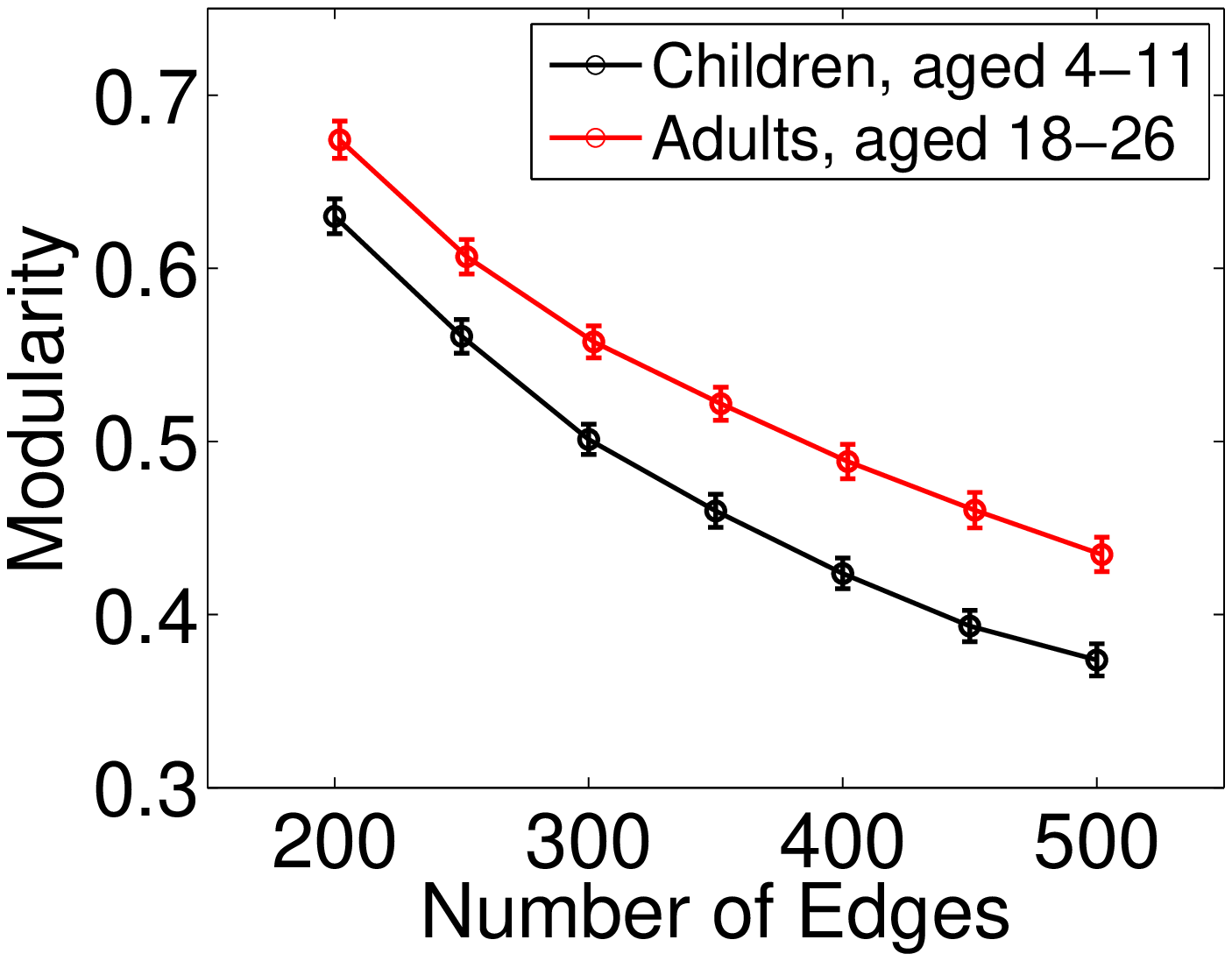}
b) \includegraphics[height=0.45\columnwidth,clip=]{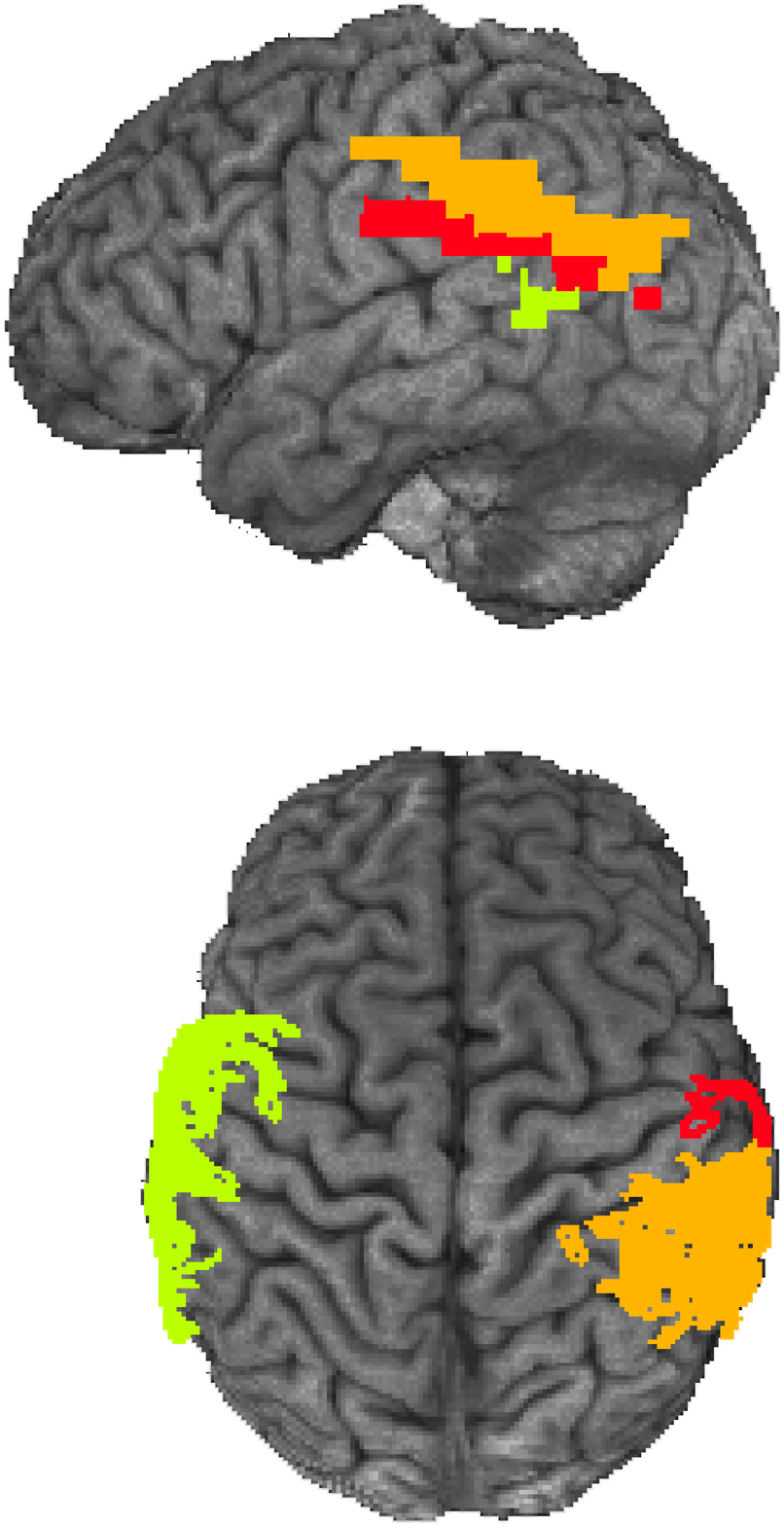}
\end{center}
\caption{a) Average modularity of the neural activity in the brain
for the child and adult cohorts.
Modularity is greater for adults than for children.
Modularity is computed from the correlation matrix of
neural activity between
Brodmann areas.  
The number of entries in the correlation matrix above the
cutoff, denoted by edges, is chosen so that the
matrix is fully connected yet still sparse.
Modularity computed using different values of the cutoff
persistently shows a higher value for adults than
for children.
The error bars are one standard error.
b) top) The three Brodmann areas whose domains grow the
most in size from children to adults, and
bottom) the three Brodmann areas whose domains shrink the most,
for 400 edges.
}
\label{fig3}
\end{figure}
Figure \ref{fig3}a shows modularity of neural activity for 
children and adults.
The modularity of neural activity of adults is greater than
that of children ($p$-value $< 0.001$ for 400 edges).
This result persists when
different cutoff values are used to calculate
modularity from the correlation matrix.
The results for 400 edges are representative, and unless otherwise
noted, we used this criterion to construct the projected correlation
matrix.
Modularity of neural activity
increases with age during childhood development.
The average Pearson correlation between modularity and age,
corresponding to the data in 
Fig.\ \ref{fig3}a, is 0.44 ($p$-value 0.02, sample size 22).
Similarly, the average Pearson
correlation between 
modularity and raw, not-age-normalized
KBIT-2 overall IQ score
is 0.234 ($p$-value 0.16, sample size 16, as some children did not
have reported IQ scores).
The positive correlation of modularity with both age
and raw IQ shows the development of
modularity in neural activity in the brain during
childhood.



\subsection{Modularity of a Full Matrix with Real-Numbered Connection Weights}

\begin{figure}[tb!]
\begin{center}
\includegraphics[width=0.9\columnwidth,clip=]{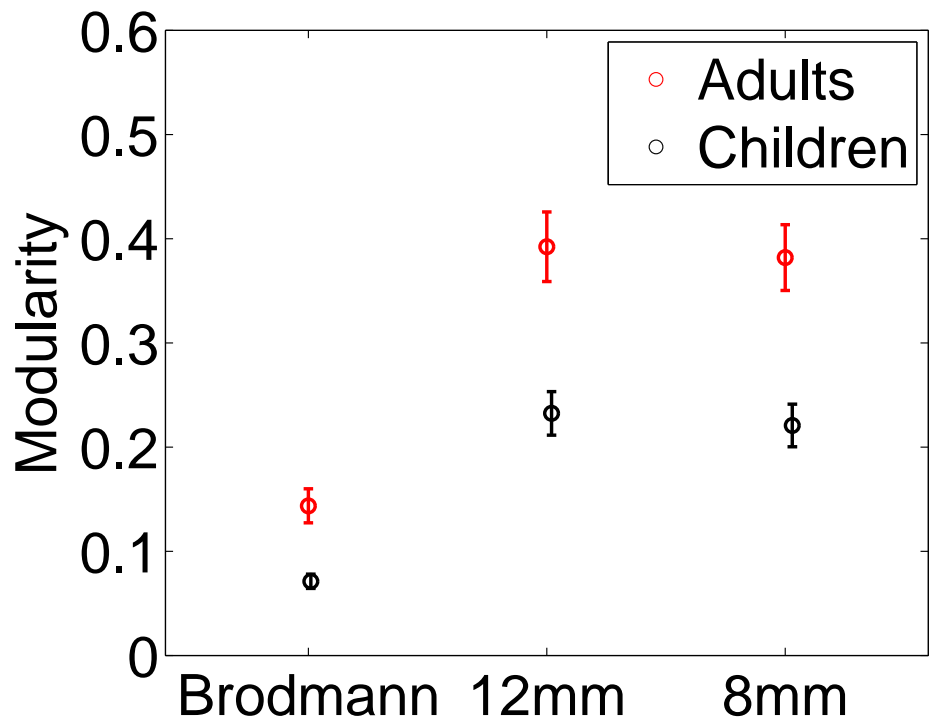}
\end{center}
\caption{Calculation of modularity when the full matrix of correlations
is used. Calculations were performed using Brodmann areas as nodes
and a $84 \times 84$ matrix of correlations.
Calculations were also performed without masking the data to Brodmann areas,
and using the original data at a resolution of 12~mm and a
$2160 \times 2160$ correlation matrix
or a resolution of 8~mm and a
$6426 \times 6426$ correlation matrix.
The $p$-values for the significance of the difference between the
modularity of adults and children are
$7 \times 10^{-5}$,
$9 \times 10^{-5}$, and
$4 \times 10^{-5}$, respectively.
 These results confirm the generality of the
results in Fig.\ \ref{fig3}.  Modularity develops during childhood.
\label{sfig0}
}
\end{figure}

Figure \ref{sfig0}
shows the results when the full matrix is computed, 
the  ``Brodmann area'' values.
 The modularity of adults and children are significantly different,
p-value $7 \times 10^{-5}$.

\subsection{Effect of Head Motion on Calculation of Modularity.}

\begin{figure}[tb!]
\begin{center}
 \includegraphics[width=0.45\columnwidth,clip=]{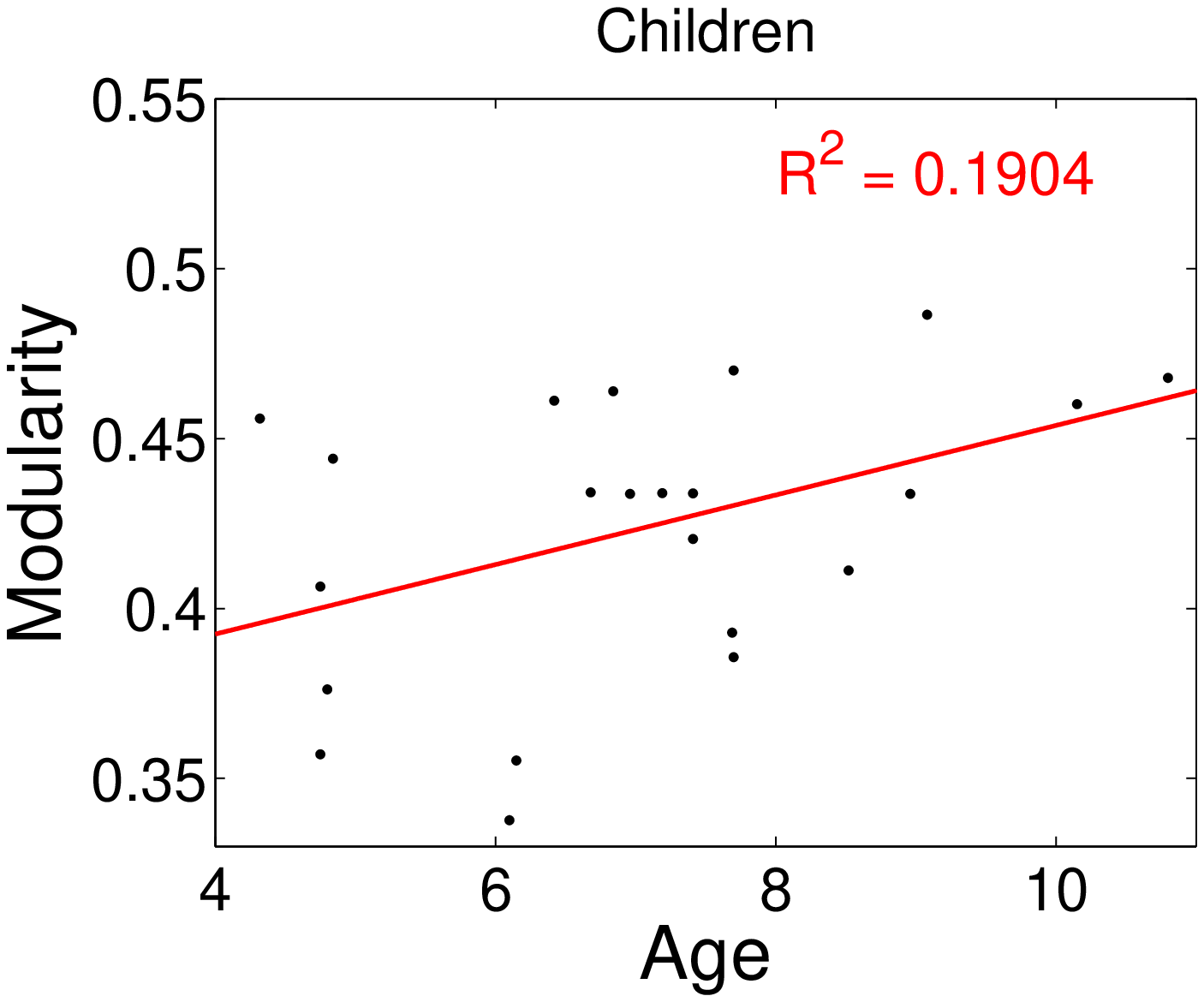}
 \includegraphics[width=0.45\columnwidth,clip=]{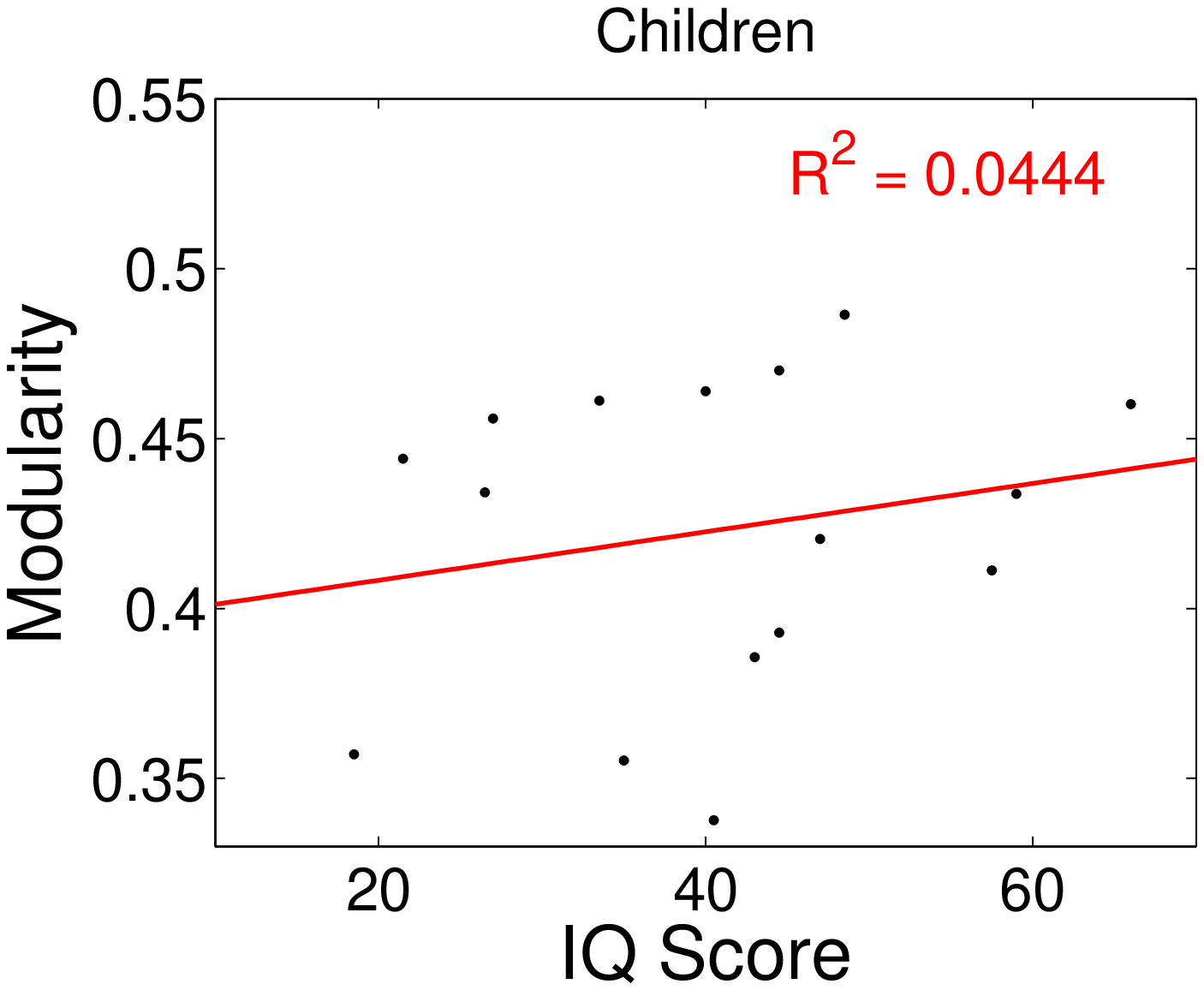}
 \includegraphics[width=0.45\columnwidth,clip=]{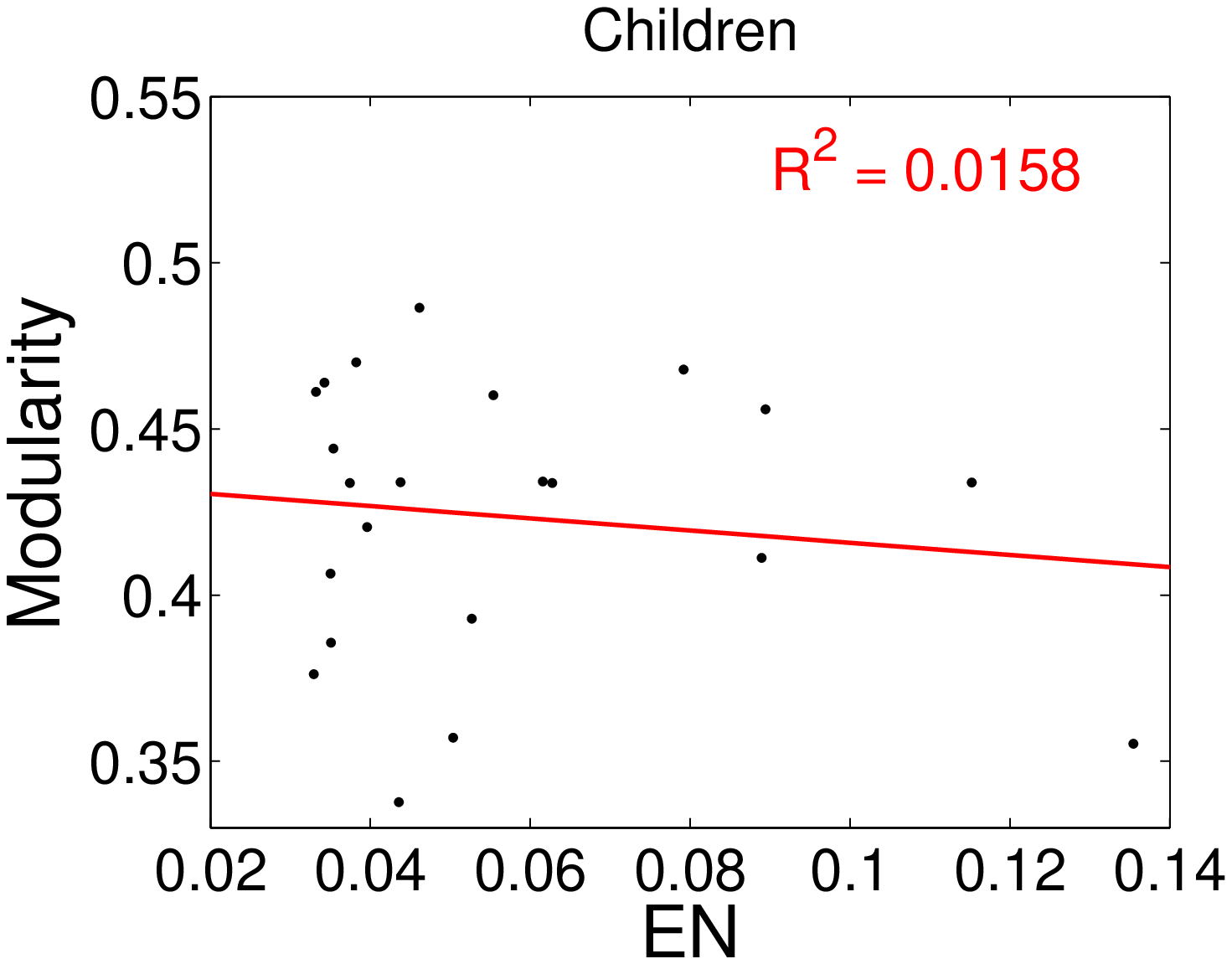}
 \includegraphics[width=0.45\columnwidth,clip=]{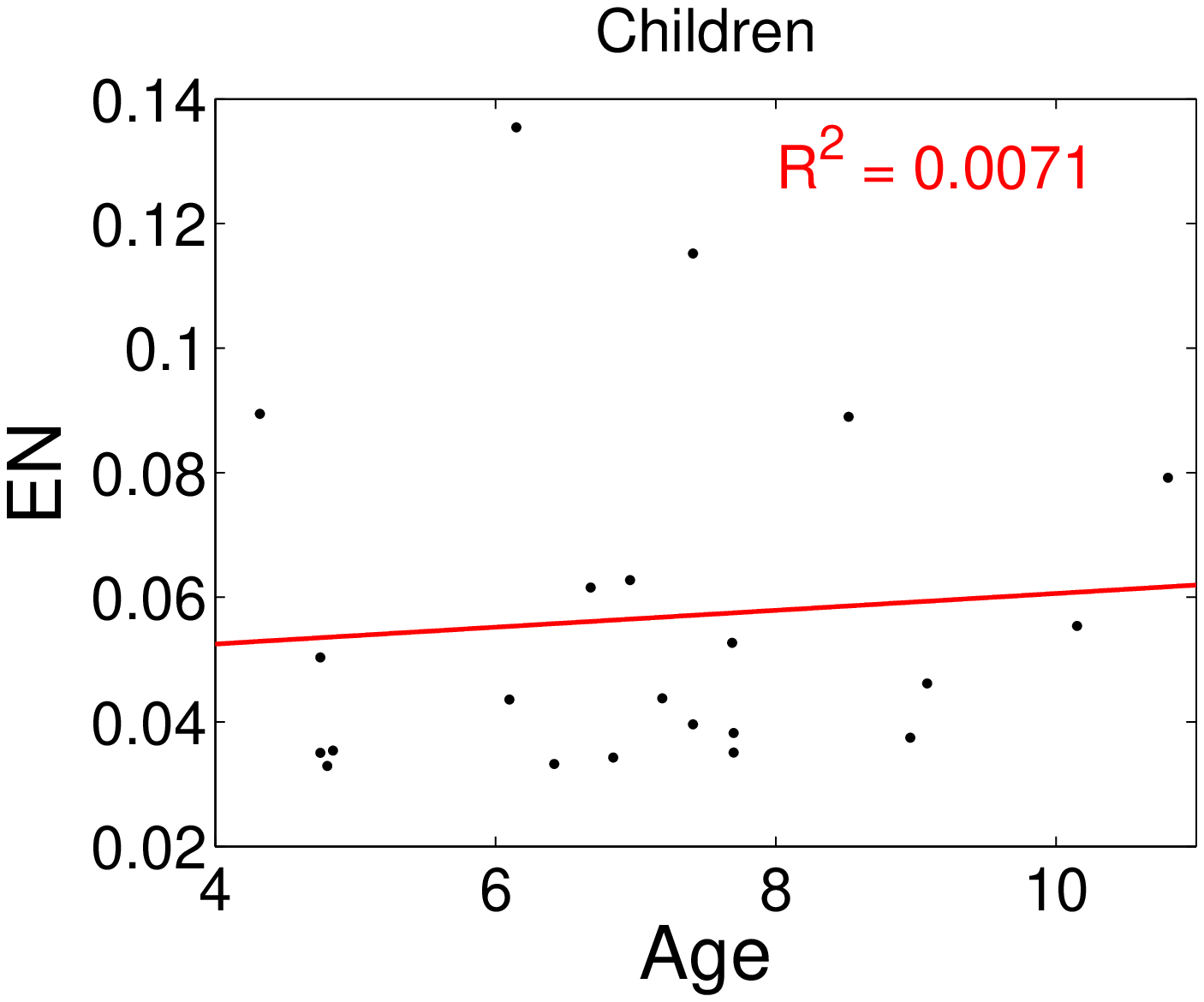}
\end{center}
\caption{The correlation of modularity with age, IQ score, and Euclidean
norm (EN) from AFNI for children.  Also shown is the correlation of EN with age.
\label{sfig2}
}
\end{figure}

The extent of head motion was measured by the Euclidean norm of the
derivatives of the motion parameters, termed EN.
  To illustrate this point, a number of correlations were
carried out to illustrate the effect of head motion. 
 The correlations for
children are shown in Table \ref{tables2} and
Fig.\ \ref{sfig2}.
\begin{table}
 \center
 \caption{Quantification of the persistence of
modularity and module identity with
optimization of modularity.}
 \begin{tabular}{c c c c c}
\hline
Dependent variable & $R^2$ & $p$-value & $\partial M / \partial$ age\\
and independent variable(s)\\
\hline 
M with age & 0.1904 &  0.0211  & 0.0102\\
M with EN &  0.0158&  0.2897 \\
EN with age &  0.0071& 0.3545 \\
M with age and EN & 0.2170  & 0.0490  & 0.0106\\
\hline
\end{tabular}
\label{tables2}
\end{table}
Modularity is significantly correlated with age.  Furthermore, inclusion of EN to the correlation of modularity with age only very slightly increases the goodness of fit ($R^2$), and does not change the positive slope of the correlation of M with age. 
The coefficient relating modularity to age,
 $\partial M / \partial$ age, is the same
whether head motion is included as an independent variable or not.
Thus, the head motion is not biasing the estimated relationship
between modularity and age.
The correlations of modularity or age with EN are small
and not significant.
The correlation of modularity with EN for adults
is also small and not significant, $p$-value = 0.32.
\begin{figure}[tb!]
\begin{center}
 \includegraphics[width=0.9\columnwidth,clip=]{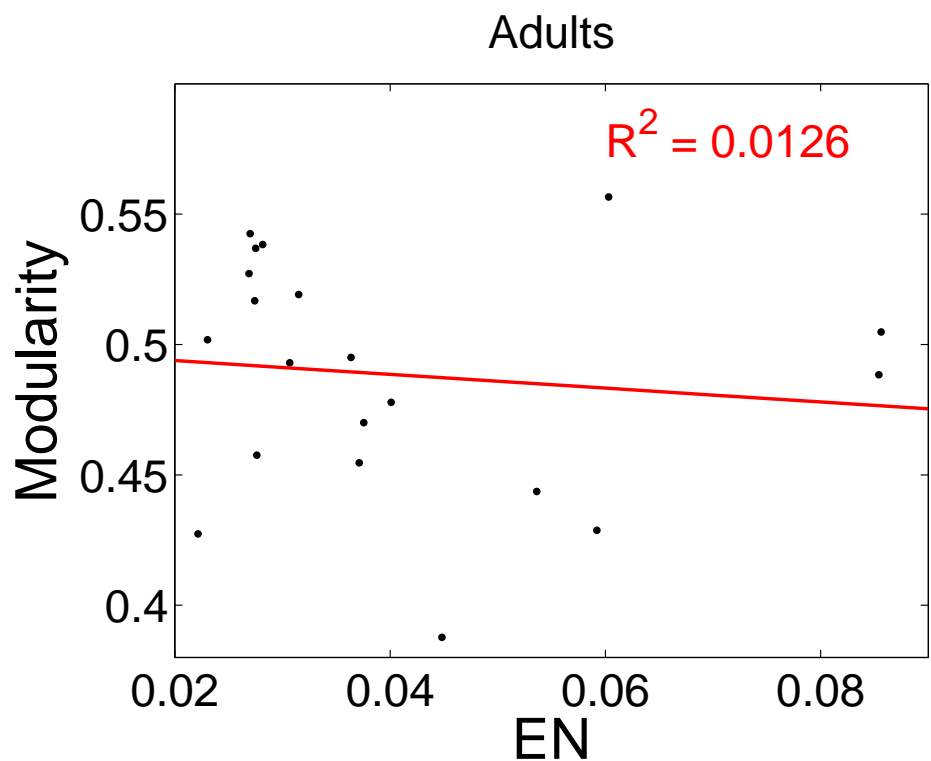}
\end{center}
\caption{The correlation of modularity with Euclidean
norm from AFNI (EN) for adults.
The correlation is small and not significant,
$p$-value = 0.32.
\label{sfig3a}
}
\end{figure}

\subsection{Development of Modules}
Not only is the modularity of neural activity in the children
and adults different, but also the identity of the modules
changes with development.  We computed the probability that
Brodmann areas $i$ and $j$ were in the same module, 
as estimated from the data by the observed fraction.
  Consider a single person.  The probability that area $i$ and $j$ are in
the same module, $p_{ij}$,
 is either 0 or 1 (they either are, or are not, in the same
module in one given subject sample).
The sum of this quantity over $j$ is the size of
the module in which $i$ is a member, $S_i$. Averaging $S_i$ 
over all children or all adults gives the average size of the
 module in which $i$ participates,
$\langle S_ i \rangle_{\rm child}$
or $\langle S_ i \rangle_{\rm adult}$, respectively.
 These two quantities were calculated
for each Brodmann area.  Also reported
are the three $i$ for which
$ \langle S_ i \rangle_{\rm adult} - \langle S_ i \rangle_{\rm child} $
is largest and the three $i$ for which
$ \langle S_ i \rangle_{\rm adult} - \langle S_ i \rangle_{\rm child} $
is smallest.  

We found the three Brodmann areas that had
the largest positive difference between
the average module size in adults and children, shown in
Fig.\ \ref{fig3}b.
 These are the areas whose modules grew most in
size with development.  
They are
left Brodmann area 23, 29, and 31.
We also found the three Brodmann areas that had the 
largest negative
difference in average module size between adults and children, i.e.
the areas whose modules shrink most with development, shown in
Fig.\ \ref{fig3}b.
They are left Brodmann area 21, 
right 40, and right 43.

\subsection{Near Degeneracy of Modularity Values}
\begin{figure}[tb!]
\begin{center}
\includegraphics[width=0.9\columnwidth,clip=]{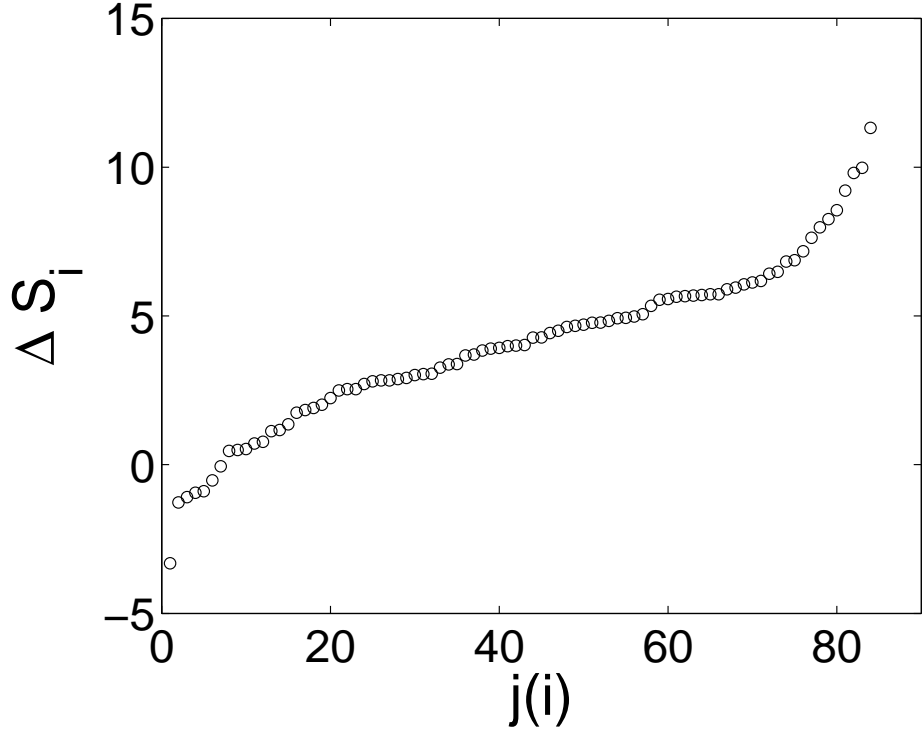}
\end{center}
\caption{The distribution of the  values of
$ \langle S_ i \rangle_{\rm adult} - \langle S_ i \rangle_{\rm child} $
They are ordered from smallest to largest; $j(i)$ denotes this
ordering.
\label{sfig0a}
}
\end{figure}

In the definition of modularity, Eq.\ (\ref{1}),
there can be partitions that give values of modularity
near but slightly below the optimal value.   The optimal
value is denoted by $M$, and the nearby values are denoted
by $Q$.
To address the near degeneracy of modularity values,
that is values of $Q$ that are near the optimal
value of modularity $M$, 
the Newman algorithm was
generalized to include the possibility of accepting a move that
decreases $Q$, with probability $\min[1, \exp(\Delta Q/T)]$.
This generalization leads to sampling the near-optimal values of Q,
roughly in the range $M-T$ to $M$.
How the average $Q$ varies with $T$ was calculated.
Also calculated was how often the set of the 3 Brodmann areas for
$ \langle S_ i \rangle_{\rm adult} - \langle S_ i \rangle_{\rm child} $
is largest changes identity in 100 runs, and similarly for
the set of the 3 for which it is smallest
.  From these results, one sees that $M$
values reported are representative, i.e. the nearly degenerate $\langle Q \rangle$ values
are close to the optimal value, $M$.  The identification
of 3 Brodmann areas for which
$ \langle S_ i \rangle_{\rm adult} - \langle S_ i \rangle_{\rm child} $
is largest and the 3 for which it is smallest
is relatively stable among these nearly
 degenerate states.    Essentially only the least stable member of the
latter is lost at finite $T$ (3/3 and 2/3 of the members
are stable, respectively).
\begin{table}
 \center
 \caption{Quantification of modularity and module identity persistence with
optimization of modularity.}
 \begin{tabular}{c c c c c}
\hline
$T$ & $\langle Q\rangle$  & $\langle Q \rangle$  & (Number top areas & (Number of bottom areas \\
&(adults) & (children) & in common with  & in common with \\
&&& $T=0$ case)/3 &  $T=0$ case)/3 \\
\hline
0 &  0.4885 & 0.4237 & 1.00 & 1.00 \\
0.01 &  0.4885 & 0.4237 & 1.00 & 0.65 \\
0.05 &  0.4883 & 0.4233 & 0.97 & 0.62 \\
0.10 &  0.4871 & 0.4220 & 0.86 & 0.60 \\
\hline
\end{tabular}
\label{tables1}
\end{table}
The top areas are more stable than the bottom areas, because there is a
gap in the distribution of
$ \langle S_ i \rangle_{\rm adult} - \langle S_ i \rangle_{\rm child} $
after the 3rd highest value, see Fig.\ \ref{sfig0a}.
This distribution shows there is a natural set of highest and lowest
outliers.

\section{Model}

\subsection{Model of the Response Function for Memory Recall}
We will explore a mechanism to understand the results with
a model of neural activity and cognitive function.
We build upon the Hopfield neural network model \citep{Hopfield}.
Models of brain activity with different levels of detail and complexity
have been developed.  At the detailed level, there are models of individuals
neuron activation and spiking \citep{FitzHugh,Nagumo,Wang}.
These models have been generalized to a population of neurons,
in which synchronization of the spike trains has been calculated
\citep{Buzsaki,Rubinov2011}.
The data analyzed are at the resolution of 4~mm, and 
a voxel of (4~mm)$^3$ contains
roughly a million neurons.  The modeling is, therefore, of
interactions between groups of neurons, with each group containing
a million neurons.  In addition, the time resolution 
of the fMRI BOLD signal data is 2 s.
The fMRI BOLD signal results from the difference in
magnetic spin properties of oxygenated and deoxygenated hemoglobin.
This signal is, therefore, a representation of blood flow to each
region of the brain.  The regional blood flow is an indication of
local neural activity \citep{Ogawa}.
 At this time and length scale, 
a model such as the  Hopfield neural network is appropriate.  As with
the  data, neural activity is the only observable in this model.

We used a neural network model to describe 
the dynamics of neural activation that was measured by the fMRI experiments.
The voxels of neural activation measured in the fMRI are 
subgroups of neurons in the brain.  In the model, the activation
state of subgroup $i$ is given by $\sigma_i(t)$,  which takes on
values $1$ or $-1$ to indicate that the
subgroup of neurons is active or inactive on average at time $t$, respectively.
Thus, for each physical region of the brain in Fig.\ \ref{fig1}a,
there is a $\sigma_i(t)$.
The neural state at time $t+1$ was created from the neural state
at time $t$, based upon connections between
neurons and stored memories. 
We here took the connections between the neurons to be modular, with
modularity $M$.
These functional connections, denoted by the matrix $A_{ij}$ below,
correspond roughly to the modules identified in Fig. \ref{fig1}d.
The model describes how neural states in the brain are
driven to match stored patterns, with the $\mu$th pattern
denoted by $\xi_i^\mu$.
We also took the stored memories to be clustered.
The correlation between these stored patterns is denoted by the
weight matrix $W_{ij}$.  The clustering of the stored memories
is quantified a parameter $p$, described below.
The modules identified in the fMRI experiments,
Fig. \ref{fig1}d, are described in detail in the model
through the combined influence of the connection matrix and the
memory correlation matrix, $A_{ij} W_{ij}$.

We related the modularity of the neural activation in the model
to the modularity of the neural activity as measured by fMRI.
The connection matrix denotes whether  neural region $i$ 
is connected to region $j$.
Due to physiological changes that occur during development,
these connections change, as shown in 
Figs.\ \ref{fig3} and \ref{sfig0}.
In particular, we set the connection matrix of the neural network
to be modular \citep{Pradhan}, with modularity $M$.
Crucially, we also set the stored memory patterns
to be clustered as well, with modularity $p$.
The neural state is propagated 
from a random initial state to recall
a stored memory according to the 
Hebbian learning rule \citep{Amit}.
We computed the average overlap between
the final neural state and the stored memory.
In terms of an experiment,
overlap at time $t$ can be interpreted as quantifying how 
well a subject correctly identifies
an image that is visible for a time, $t$.

In the neural network model, memory $\mu$ is defined by the pattern
$\xi_i^\mu=\pm 1$ of length $N=256$ bits \citep{Hopfield}.
The weight matrix is defined by $W_{ij} = \sum_\mu 
\xi_i^\mu
\xi_j^\mu$.
The $N/4 \times N/4$ block diagonals define the
four modules.  Four distinct patterns are stored, one
per module.  Each pattern $\mu$ has $N/4$ values with $\xi_i^\mu=+1$,
each of
which has probability $(1 + 3 p)/4$ of being within the $\mu^{\rm th}$
block diagonal.
The neural state is defined by $\sigma_i(t)$ and updated
by the Hebbian learning rule \citep{Amit}
$\sigma_i(t+1) = {\rm sign}[ \sum_j A_{ij} W_{ij} \sigma_j(t)]$.
The connection matrix $A$
is binary and sparse, with average degree $\langle k \rangle = 30$.
The probability for a connection to be 
at a given site within the block diagonal 
is given by $(1 + 3M) \langle k \rangle/N$.
  The overlap of the neural state
with the target memory is given by 
$\max_\mu \sum_i \xi_i^\mu \sigma_i(t)/N$.

\subsection{A Hierarchical Generalization}
\begin{figure}[tb!]
\begin{center}
\includegraphics[width=0.9\columnwidth,clip=]{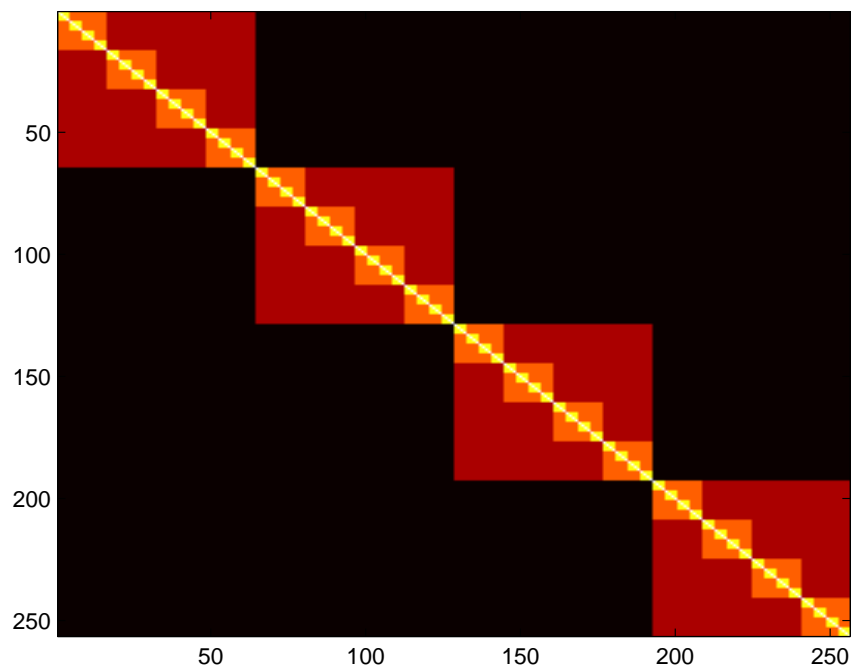}
\end{center}
\caption{Pictorial representation of the hierarchical matrix.  The
levels are 4 (black), 3 (red), 2 (orange), 1 (yellow), and 0 (white).
matrix.
\label{sfig3b}
}
\end{figure}

It has been argued that at the largest scales, the brain structure is
hierarchical, not simply modular.  Thus, the modular
Hopfield neural network described above
was also generalized to an hierarchical model.
 A similar hierarchical model was studied by Rubinov \emph{et al}.\
in a computational analysis
to show the effect of self-organized criticality and
neuronal information processing \citep{Rubinov2011}.
  In the modular model, the connection matrix $A$ is made modular.
In the hierarchical model discussed here,
5 levels, $\gamma$, are defined.  The level for matrix position $ij$ is defined
as the smallest $\gamma$ for which
 $\lfloor i/4^\gamma \rfloor 
=
\lfloor j/4^\gamma \rfloor $.
For example, $\gamma = 0$ is the diagonal, and $\gamma = 4$ is the
entire matrix, excluding the all the $4 \times 4$ block diagonals.
See Fig.\ \ref{sfig3b}.
The probability to be within region $\gamma$  is assigned to be
proportional to $1 - \gamma \epsilon $, with the proportionality
determined so that the average number of connections
per node  remains at 30.
Here $\epsilon= 1/4$ is a  measure of the asymmetry introduced by
the hierarchy.

\subsection{Model Results}

\begin{figure}[t!]
\begin{center}
\includegraphics[width=0.95\columnwidth,clip=]{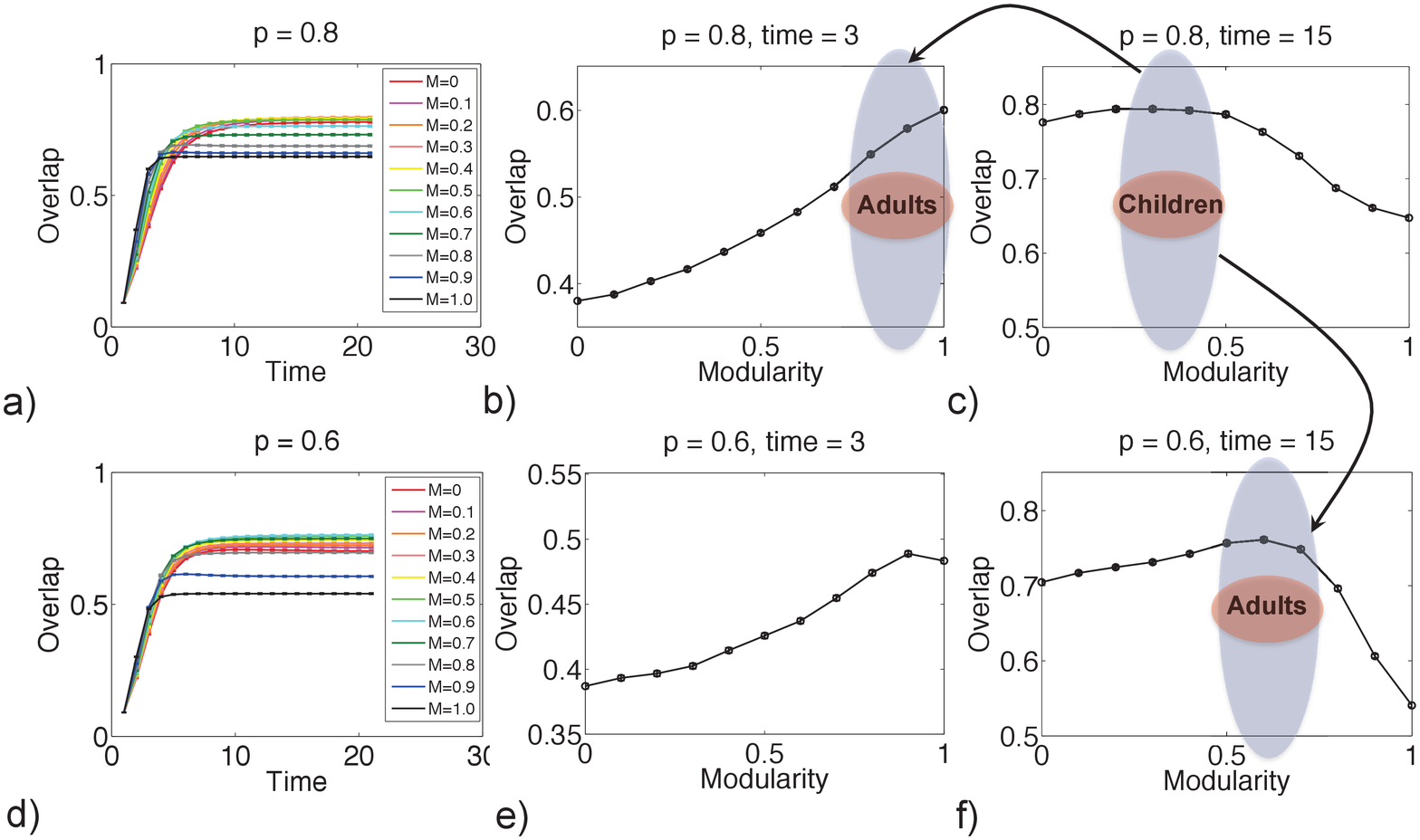}
\end{center}
\caption{
More modular neural architectures give better 
performance at short times (a or d, short time) and less modular
memory architectures can give better performance at long times (a or d, 
long time).
The greater modularity in adults than children, Fig. \ref{fig3},
is consistent with either cognitive performance
at short times is more important in adults than children (top arrow)
or that memories are less clustered in adults than children (bottom arrow).
Overlap is a measure of the probability that the neural state
correctly recalls a memory.  The modularity of the connection
matrix 
is $M$.  The timescale is of order seconds.
The clustering of the stored patterns is denoted by $p$.
}
\label{fig2}
\end{figure}
The cognitive ability of the brain, i.e.\ its
ability to solve a challenge,
depends on modularity of the neural activity.
The responses of the model neural system with high modularity
and lower modularity are shown in 
Fig.\ \ref{fig2}.
In this figure, cognitive performance is quantified by
overlap between evolved neural state and target memory state.
These performance curves as a function of time and 
neural architecture 
depend on the clustering of the stored memories, $p$.
The overlap typically increases with time, as the target
memory is dynamically recalled.  At short times, a
more modular memory
architecture can lead to a better recall, i.e.\ greater overlap
with the stored memory.  At longer times, a less 
modular memory architecture can give better performance.
We view this crossing of the performance as a function of time
to be a generic result of an evolving dynamical system with
a rugged fitness landscape \citep{Deem2013b} and not
unique to the particular model used here.

The variation in performance, shown in Fig.\ \ref{fig2},
helps to explain the  change
of modularity during development.
We use quasispecies theory to quantify the relationship between
performance and change of modularity.
In this theory, systems with different modularity, $m$, are assigned
a fitness, $f(m)$, that quantifies the benefit of performance.
This theory predicts how modularity changes with time,
given the fitness function, and the rate at which entries
in the connection matrix change, $\mu$.

\begin{figure}[t!]
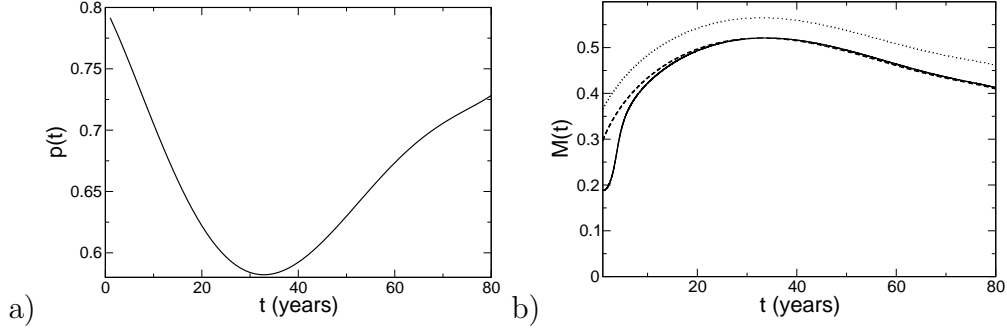

\begin{center}
a) \includegraphics[width=0.4\columnwidth,clip=]{Fig5a.eps}
b) \includegraphics[width=0.4\columnwidth,clip=]{Fig5b.eps}
\end{center}
\caption{a) The 
clustering of memories versus age, after
the bottom arrow in 
Fig.\ \ref{fig2}.
b) The average modularity versus age
predicted by quasispecies theory (solid).  Here, the fitness
is $10 \times$ the overlap in
Fig.\ \ref{fig2}, and the
rate of mutation is $\mu = 0.1$
\citep{Deem2014}.
Also shown is the adiabatic approximation to the modularity,
$M^\infty$ (dashed)
as well as the modularity that maximizes the fitness,
$M^*$
(dotted).
}
\label{fig5}
\end{figure}

We take the overlap in Fig.\ \ref{fig2} 
as the fitness for modularity, $f(m)$,  in the brain and use
quasispecies theory 
\citep{Deem2014}
to predict how modularity develops with age.
The average modularity, $M(t) = \langle m(t) \rangle$, changes with age
according to
\begin{equation}
\frac{d M}{d t} = N \langle m f[p(t),m] \rangle - N M \langle f \rangle - \mu M
\label{eq1}
\end{equation}
where $\mu$ is the rate of  mutations in the connection matrix
\citep{Deem2014}.
Following the bottom arrow of Fig.\ \ref{fig2},  
the response time is taken to be 15, and it is assumed
that $p(t)$ changes from 0.8 at birth to 0.6 in middle
age and 0.7 in old age, 
This $p(t)$ is shown in 
Fig.\ \ref{fig5}a. 
Previous studies have shown that the modularity, measured
with an automated anatomical labeling (AAL)
basis set for resting state activity, of
neural activity in
adults with average age of 70 is roughly 7\% 
below that of
adults with average age of 22 \citep{Meunier2009,Dirk}.
Another study of 193 adults aged 34--87, also using the AAL basis
set, found that on average, modularity 
decreased 8\% over 50 years \citep{Onoda}.
These results imply $p$ should be lower in old age than in young adulthood.
The prediction from Eq.\ (\ref{eq1}),
shown in Fig.\ \ref{fig5}b,
is a qualitative prediction for the developing
human brain.
Due to mutation \citep{Deem2014}, the observed modularity 
in Fig.\ \ref{fig5}b is below the value that
maximizes the fitness, $M^*(t)$, defined by
$f[p(t), M^*(t)] \ge f[p(t), M] ~\forall~ M$.
We calculate the modularity in the adiabatic limit,
$M^\infty(t)$,
from the steady-state average fitness 
derived from a solution of the quasispecies theory \citep{Deem2014}:
\begin{eqnarray}
f_{\rm pop} &=& \max_\xi \{ f[p(t), \xi] - \mu C [(N-L) L / N^2]
[2 + (N/L-2) \xi  
\nonumber \\ && - 2 \sqrt{
(1-\xi) (1 + (N/L-1) \xi)
}
]
\}
\nonumber \\
&=& f[p(t), M^\infty(t)]
\label{eq2}
\end{eqnarray}
where $C$ is the average number of connections in the
connection matrix, per row.
%
%

\begin{figure}[tb!]
\begin{center}
a) \includegraphics[width=0.25\columnwidth,clip=]{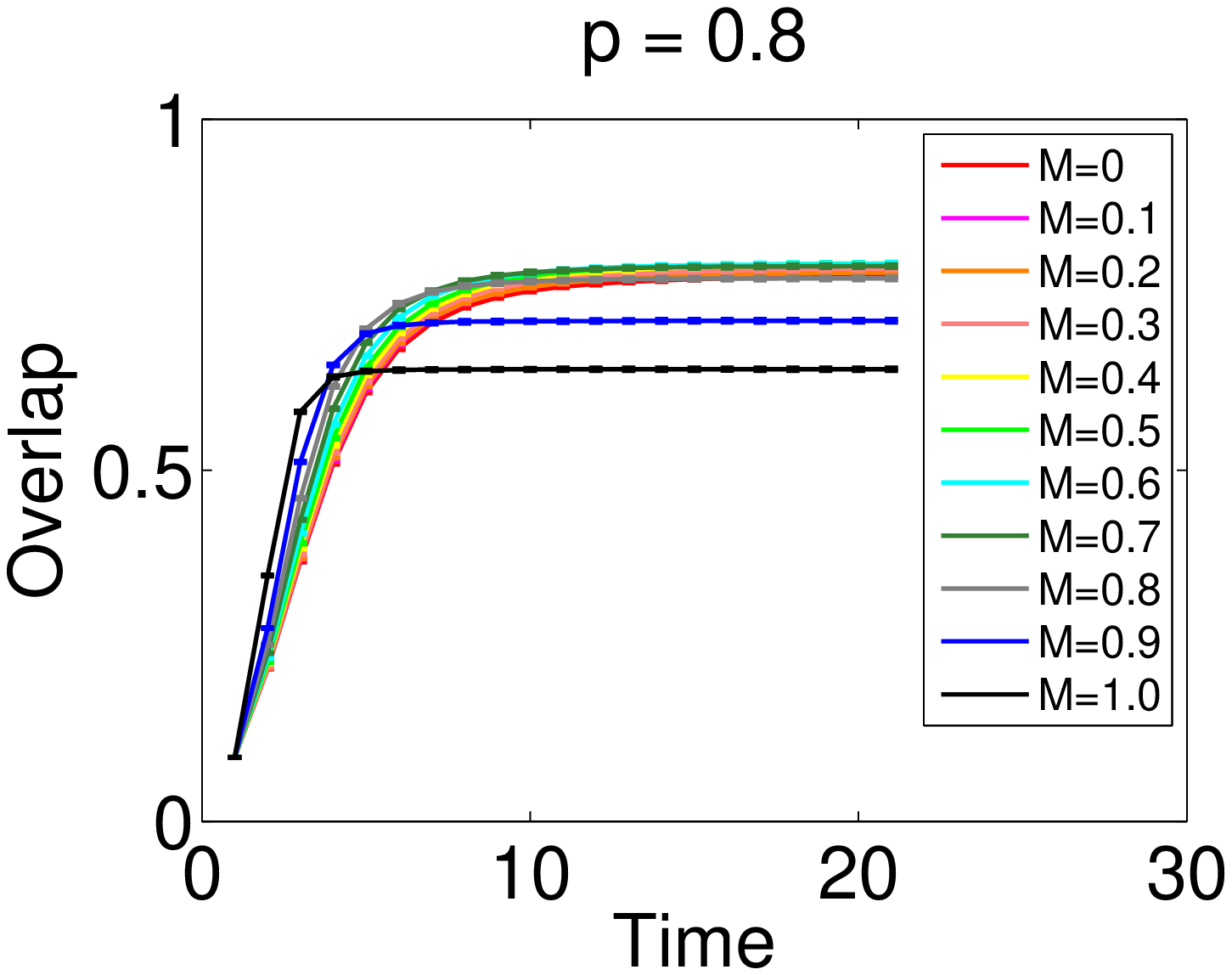}
b) \includegraphics[width=0.25\columnwidth,clip=]{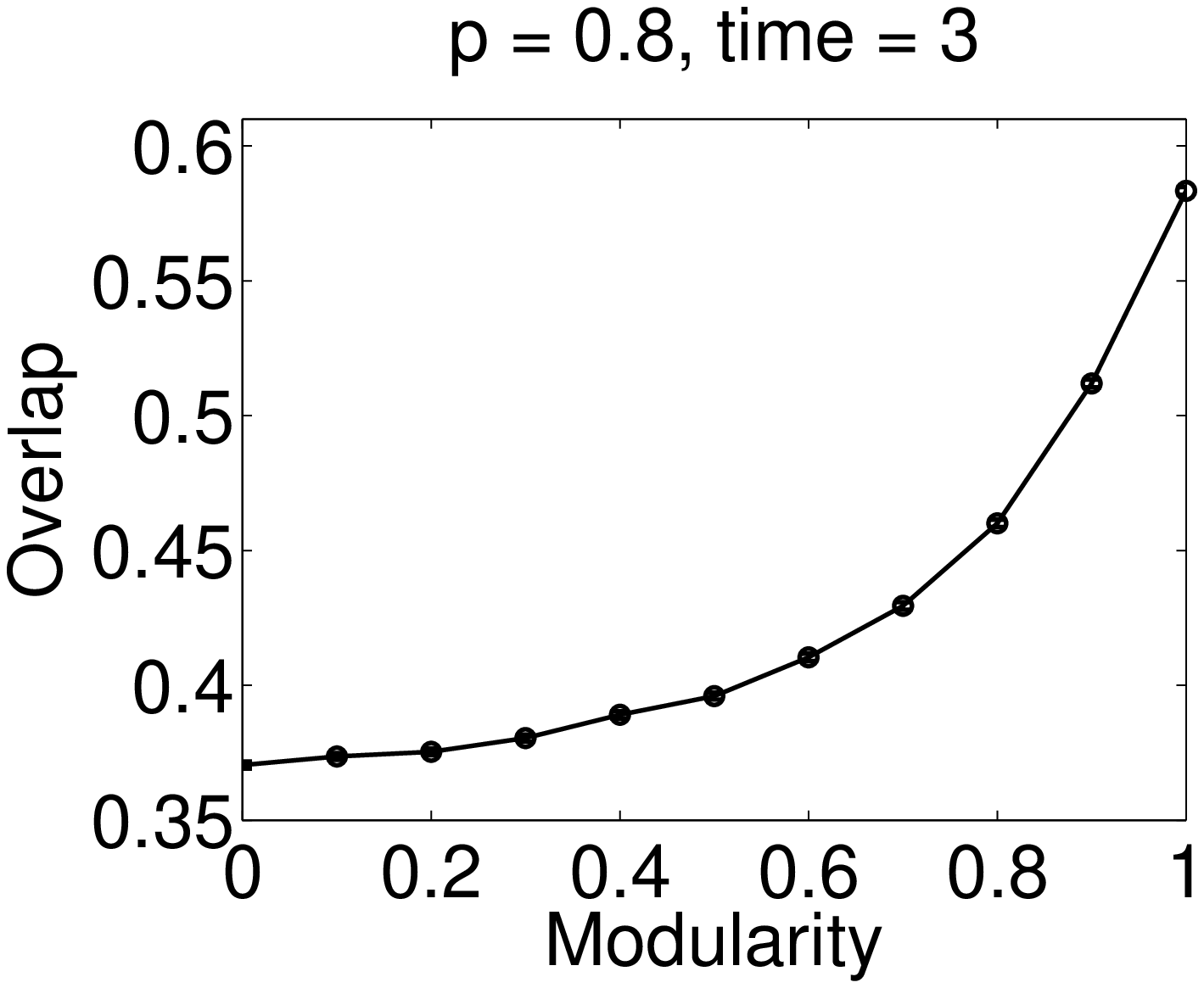}
c) \includegraphics[width=0.25\columnwidth,clip=]{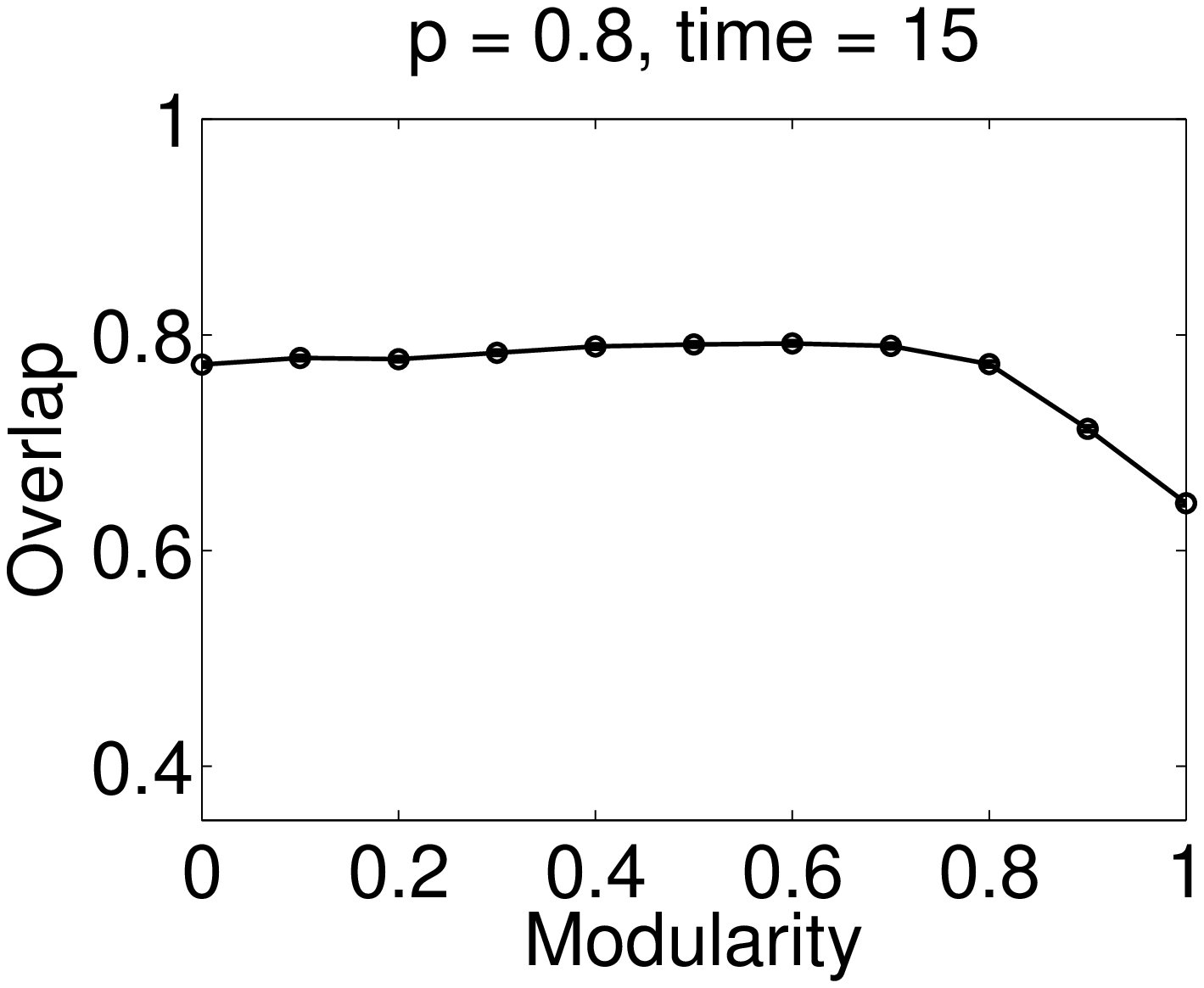}
\end{center}
\caption{Hierarchical Hopfield model results.
  Here $p = 0.8$.
\label{sfig3}
}
\end{figure}

\begin{figure}[tb!]
\begin{center}
a) \includegraphics[width=0.25\columnwidth,clip=]{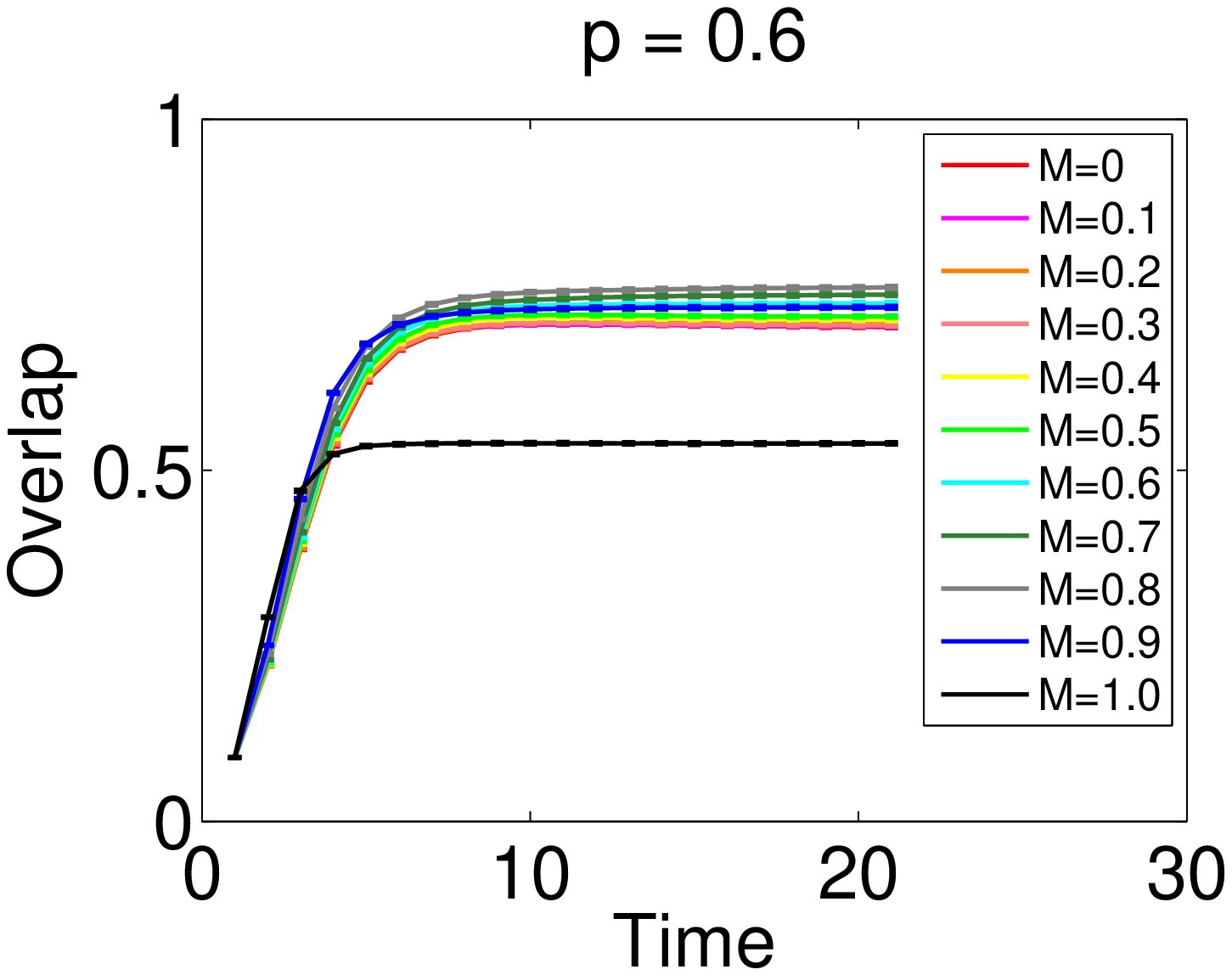}
b) \includegraphics[width=0.25\columnwidth,clip=]{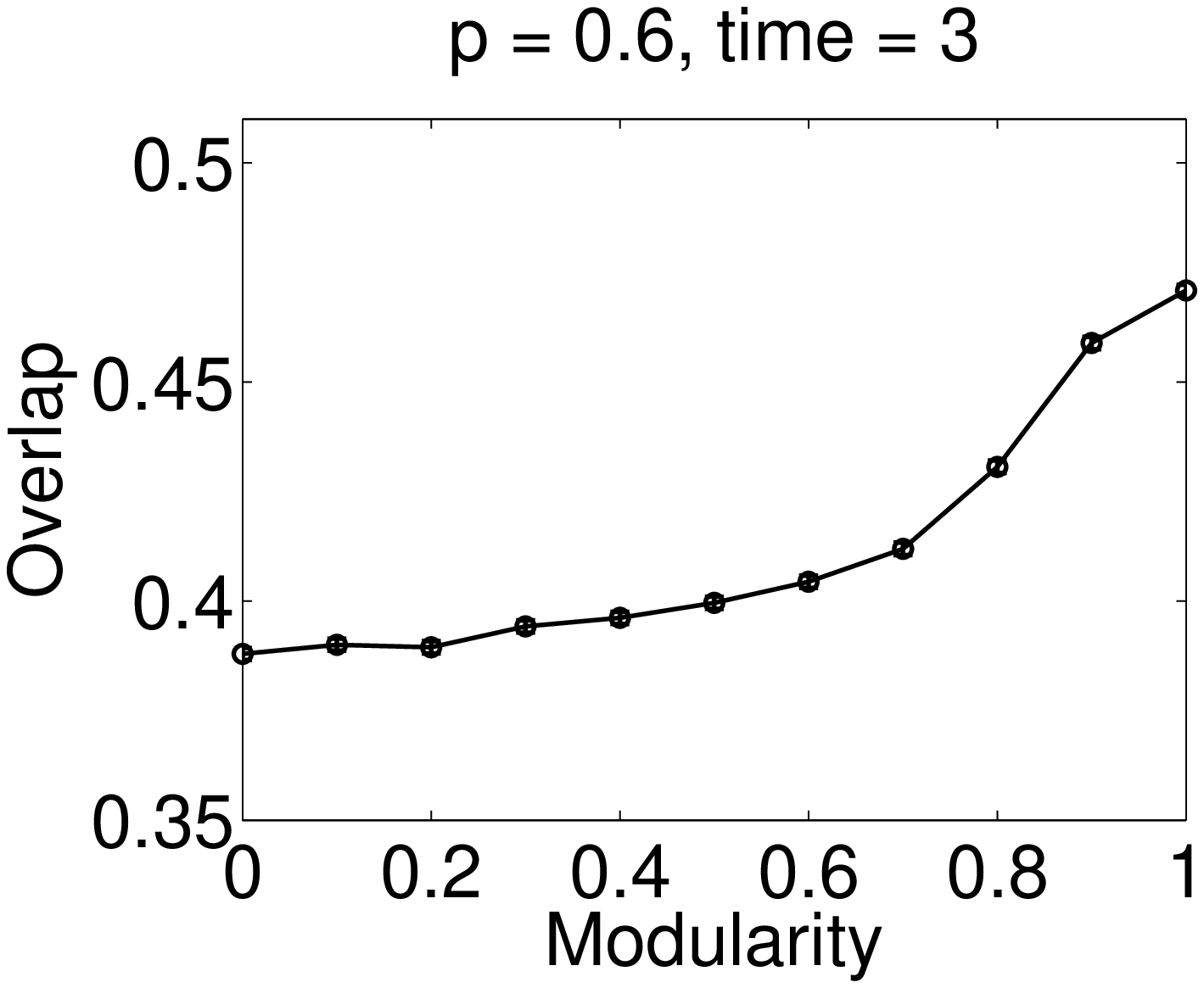}
c) \includegraphics[width=0.25\columnwidth,clip=]{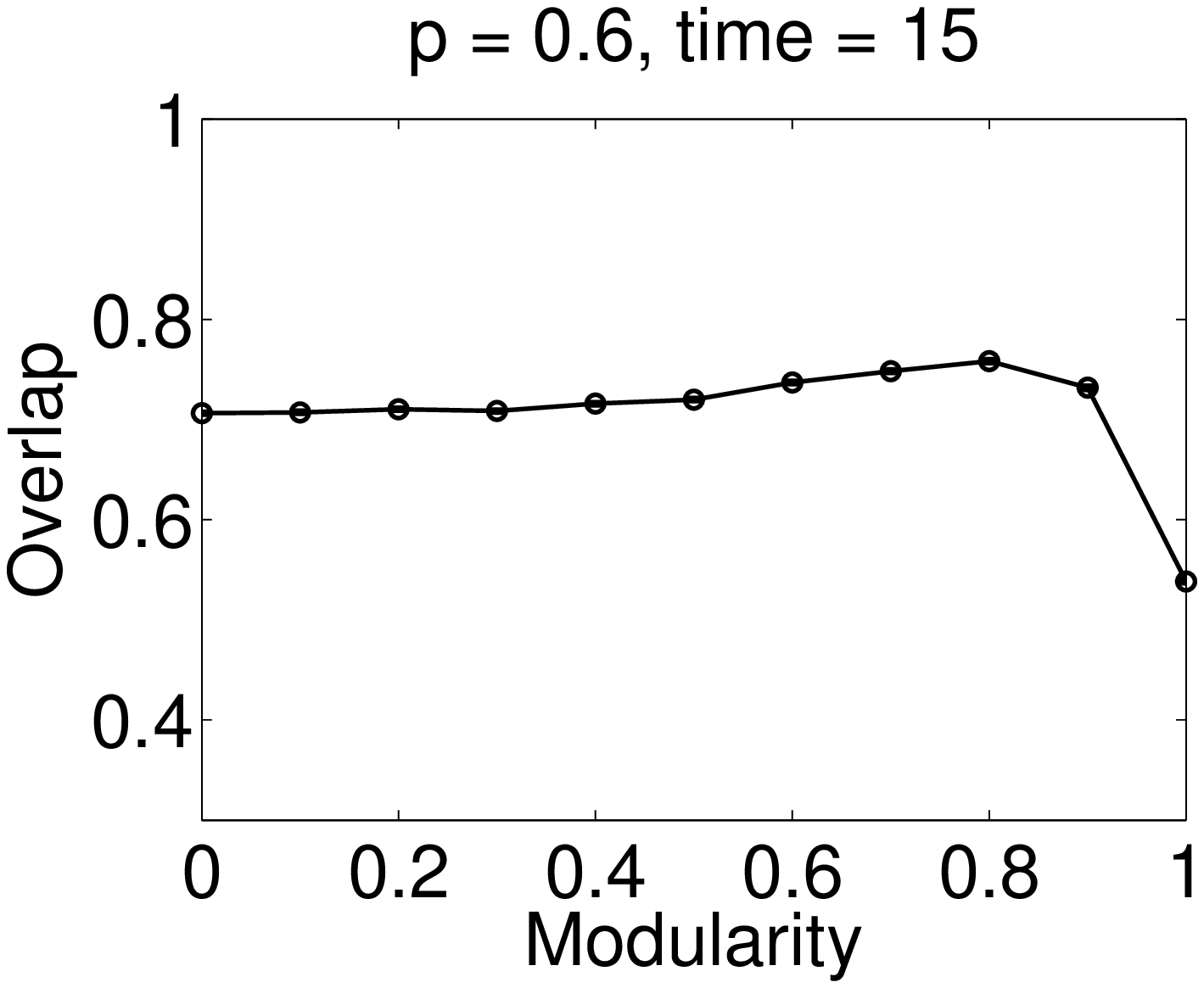}
\end{center}
\caption{Hierarchical Hopfield model results.
  Here $p = 0.6$.
\label{sfig4}
}
\end{figure}

Results for the hierarchical model,
analogous to those in Fig.\ \ref{fig3}, are shown in
Figs.\
 \ref{sfig3}
 and
\ref{sfig4}.
These results show that this hierarchical generalization of the
Hopfield model also shows a crossing of the response function at
intermediate times.  That is, low levels of hierarchy lead
to better responses at long times.  And high levels of hierarchy can
lead to better responses at short times.


\section{Discussion}

We have observed that modularity of neural 
activity in the brain increases with age from
children to young adults, as measured by fMRI experiments.
These data are summarized in Fig.\ \ref{fig3}.
The analysis separately takes into account head motion, and
the present study finds that
after censoring and alignment,
the effects of head motion are negligible.

The $N_{\rm edge}$ cutoff parameter from Fig.\ \ref{fig3}
can be viewed as a clustering parameter.  What Fig.\ \ref{fig3}
 shows is that the modularity of adults is greater than that of children, persistently with the value of the $N_{\rm edge}$ cutoff parameter.  In other words, the conclusion that modularity
develops from children to adults is robust with respect to the particular value of the cutoff parameter.

Previous analysis of fMRI data
from young adults and old adults has shown
that modularity of neural activity in the brain
decreases with age \citep{Meunier2009,Dirk,Onoda}.
Taken together with the present results, it appears
that modularity of neural activity peaks in young
adulthood.

The calculation of modularity was done using both the Brodmann
areas as a basis and using raw voxel data.  Both calculations show
the modularity of children is greater than that of young adults.
Additionally, the calculated values of modularity were shown to
be representative of the full distribution of near-optimal values.

The Brodmann areas that grew most during development,
left Brodmann area 23, 29, and 31, are in
the posteromedial cortex.  They play a central role in the brain
neural network and in communication with the rest of the brain
\citep{man1,man2,man3}.  These areas also 
play important roles in memory retrieval \citep{man4,man5}.
It is interesting that all three areas in the left brain,
generally associated with  logic, language and analytical thinking.
The Brodmann areas identified to be in the modules that grow the most
with development are somewhat sensitive to the number of edges used
in the projection of the correlation matrix.  For the range of edges
show in Fig.\ \ref{fig3}a, 80\% of the identified Brodmann areas
are in the left hemisphere.  
Additionally, the area with the
most dominantly growing
module, left Brodmann 29, is identified 80\% of the time.
Our results are consistent with the observation that
the right brain is dominant in infants,
and that left brain develops later into adulthood \citep{Chiron}.
The Brodmann areas that shank most during development,
They are left Brodmann area 21, right 40, and right 43,
are related to language perception and processing, accessing word meaning, and
face recognition \citep{man6,man7}.  The area with the
most dominantly shrinking 
module, left Brodmann 21, is identified 80\% of the time.

Prior results show that young adults
are able to more quickly solve task-switching challenges
than are children or old adults \citep{Karbach}.
The present model shows that modularity allows for rapid responses.
That is, a more modular neural activity
can allow the brain to switch more quickly 
from one type of neural activation to another.
This conclusion is robust to refinements in the hierarchical
structure of the model.
Therefore, selective pressure for rapid cognition
should lead to the emergence of modularity 
in neural activity of the brain
during childhood development.

The present model suggests that modularity of neural
activity may develop to facilitate 
rapid cognitive function.
Modularity may be larger in young adults than in
children because
the typical required response time is shorter (upper
 arrow of Fig.\ \ref{fig2}).
Modularity may also be larger because
there are more connections between memories in adults than children,
i.e.\ memories are less clustered, quantified by a smaller $p$
(lower arrow Fig.\ \ref{fig2}).
A module offers a pre-computed
solution to a problem that has been previously encountered.
Development of modularity from children to adults, thus, can improve
task-switching performance.

\subsection{Experimental Implications}

Experiments to determine
cognitive performance, as it depends on modularity,
would be interesting to carry out.
Predictions from quasispecies theory using this fitness function
could test the significance of the measured task for developmental
selection.
Cognitive performance
of subjects could be challenged 
while fMRI data are collected. 
For example, as suggested by the model, the probability of a subject
to correctly identify an image  visible for a time $t$
may be measured.
The cognitive performance should depend, among
other parameters, 
on the modularity of the correlations in the
subject's neural activity in the brain.
Measuring performance as a function of modularity would
provide the cognitive performance function,
e.g.\ Fig.\ \ref{fig2}a or d, for this particular task.
Perhaps the cognitive performance function will peak at
different values of modularity for
different tasks.  It has been suggested that cognitive processes
which are fast
are more modular than
those which are slow
 \citep{Meunier2010}.
The results in Fig.\ \ref{fig2} show
why this is the case: modular
networks provide better performance at short times
(Fig.\ \ref{fig2}b or e),
 but less modular 
networks can provide better performance at long times
(Fig.\ \ref{fig2}c or f).
Measurements of cognitive performance for 
high-level and low-level tasks would complement these
results.
We predict that performance curves as a function of time
will cross for subject samples with different values of modularity,
as in Fig.\ \ref{fig2}a.

\subsection{Modularity as a Biomarker}

A biomarker for brain function may be developed from modularity.
For example, modularity of neural activity in epileptic patients is 
less than that in normal subjects \citep{Chavez2010}.
Anecdotal evidence \citep{Leclercq} 
suggests that neural activity in patients with
traumatic brain injury (TBI) is less modular than that in
healthy subjects.
Thus, we predict that background neural activity  in the brains of
TBI patients will be less modular
than that of healthy subjects.
If so, modularity may be useful to quantify the extent of TBI,
which is currently difficult to determine.
Effectiveness of treatment is also difficult to quantify, and
measurements of modularity may be helpful to
track progress of TBI rehabilitation treatments.
Measurements of modularity may even
be useful as feedback during treatment.
Interestingly, modularity seems to increase in response to
disease progression and reduced cognitive function
in multiple sclerosis patients \citep{Gamboa},
perhaps because the system is compensating for
increased stress due to reduced function with increased modularity
\citep{Dirk}.

\section*{Acknowledgments}

We thank Jessica Cantlon for providing the
fMRI, age, and IQ data \citep{Cantlon}.  
This research was supported
by the US National Institutes of Health under grant
1 R01 GM 100468--01.

\bibliography{cognitive2}
\end{document}